\documentclass[aps,prl,reprint,superscriptaddress,nofootinbib,floatfix]{revtex4-2}

\usepackage{amsfonts,amssymb,amsmath}
\usepackage[dvipsnames, table]{xcolor}
\usepackage{epsfig}
\usepackage{bm}
\usepackage{latexsym}
\usepackage{natbib}
\usepackage{url}
\usepackage{dcolumn}
\usepackage{color}
\usepackage{graphicx}
\usepackage{booktabs}
\usepackage{tikz}
\usetikzlibrary{decorations}
\usepackage{mathrsfs}
\usepackage{hyperref}
\hypersetup{colorlinks=true,
            urlcolor=blue,
            citecolor=blue,
            linkcolor=blue,
            menucolor=blue,
            anchorcolor=blue,
            filecolor=blue}

\begin{document}

\title{Ring Spacing from a Fourth-Order Radial Feedback Green Function\\
in a Keplerian Accretion Disk}

\newcommand{\apjl}{\textit{Astrophys J Lett}}
\newcommand{\apjs}{\textit{Astrophys J Suppl}}
\newcommand{\aap}{\textit{Astron Astrophys}}
\newcommand{\aaps}{\textit{Astron Astrophys Suppl}}
\newcommand{\aj}{\textit{Astron J}}
\newcommand{\araa}{\textit{Annu Rev Astron Astrophys}}
\newcommand{\mnras}{\textit{Mon Not R Astron Soc}}
\newcommand{\pasp}{\textit{Publ Astron Soc Pac}}
\newcommand{\pasj}{\textit{Publ Astron Soc Jpn}}
\newcommand{\jcap}{\textit{JCAP}}
\newcommand{\jhep}{\textit{JHEP}}
\newcommand{\aplett}{\textit{Astrophys Lett}}

\newcommand{\TDLI}{\affiliation{Tsung-Dao Lee Institute, Shanghai Jiao Tong University, Shanghai 201210, China}}
\newcommand{\SJTU}{\affiliation{School of Physics and Astronomy, Shanghai Jiao Tong University, Shanghai 200240, China}}
\newcommand{\KEYN}{\affiliation{Key Laboratory of Modern Astronomy and Astrophysics, Nanjing University, Ministry of Education, Nanjing, China}}
\newcommand{\NJUS}{\affiliation{Institute of Science and Technology for Deep Space Exploration, Nanjing University, Suzhou, China}}

\author{Yiwei Bao}
\email{sjtu0538015@sjtu.edu.cn}
\TDLI \SJTU

\author{Can Cui}
\email{ccui@nju.edu.cn}
\NJUS

\begin{abstract}
ALMA observations of protoplanetary disks reveal ubiquitous concentric ring
structures whose origin remains debated.  We present an exactly solvable local
model for ring spacing in a Keplerian disk.  The passive advection--diffusion
problem with a general vertical diffusivity profile separates into a vertical
Sturm--Liouville spectrum and a radial modified-Bessel Green function.  This
passive kernel is smooth and does not by itself generate a periodic ring train.
We therefore introduce a minimal fourth-order radial feedback closure for the
ring-averaged surface density.  For a localized steady ring source, the
resulting Green function has a damped oscillatory exterior branch whenever the
decaying spatial roots are complex.  Under an observationally motivated AU
scaling, the same dimensionless solution gives an illustrative spacing of order
$12$~AU, distinct from the fastest-growing temporal wavelength.  The model
separates the passive transport kernel from the feedback mechanism that selects
the observable radial scale.  Because this mechanism is internal to the disk,
it does not require pre-existing planets.  The vertical diffusivity profile
affects the background kernel but is not the source of the Green-function
oscillation.
\end{abstract}

\maketitle

{\bf\emph{Introduction.}}---
The high-resolution ALMA image of HL~Tau \cite{ALMA2015} made ring--gap
structure a central problem in planet formation.  The disk contains multiple
bright concentric rings between $\sim$10 and $\sim$80~AU, with spacings of
order 10--20~AU and contrasts large enough to rule out a smooth power-law
continuum model.  The DSHARP survey subsequently showed that such
substructures are common: nearly every sufficiently resolved disk displays
rings, spirals, or crescents \cite{Andrews2018}.  Systems including TW~Hya
\cite{Andrews2016}, HD~163296 \cite{Isella2016}, and AS~209 \cite{Fedele2018}
show multiple axisymmetric rings with characteristic separations of
20--40~AU \cite{Huang2018}.  These observations motivate a basic theoretical
question: what sets a radial length scale for repeated rings?

Embedded planets are the most developed explanation.  Planet--disk
interactions can open gaps, trap dust near pressure maxima, and generate
ring--gap pairs \cite{Dong2015,Zhu2015}.  However, many ringed disks have no
directly detected planet, and disks with many rings may require several
perturbers.  Moreover, annular substructures are observed in very young
embedded disks, including a Class~I disk younger than $5\times10^5$~yr
\cite{SeguraCox2020}, when forming planetary cores at tens of AU remains
challenging \cite{Lambrechts2014}.  A mechanism that does not require
pre-existing planets could therefore operate before planet formation is
complete.  Internal disk processes provide another route: dead-zone transitions,
snow lines, zonal flows, photoevaporative structures, and dust evolution can
all create axisymmetric substructure
\cite{Gammie1996,Hasegawa2010,Flock2015,Zhang2015,Johansen2009,Bai2014,
Ercolano2017,Pinilla2017}.  These mechanisms differ in their predicted
spacings, widths, and contrasts.  An analytic model is therefore useful even
when it is phenomenological, because it isolates which part of the physics
selects the observable radial scale.

Here we develop such a model for a Keplerian disk with a passive vertical
transport kernel.  For a broad class of positive vertical diffusivity profiles,
the axisymmetric passive problem separates into a Sturm--Liouville eigenvalue
problem in $z$ coupled to a modified Bessel Green function in radius.  The
passive calculation is exact and important: it gives the transport kernel on
which any ring-forming mechanism must act.  It does not, by itself, produce a
robust periodic train of rings.  The complete positive passive response is
dominated by the lowest modes and is a smooth transport kernel.

The scale selection in this paper comes from a second ingredient: a local
fourth-order radial feedback closure for the ring-averaged surface density.
The closure represents an effective response of drift and transport to local
radial gradients, with a fourth derivative regularizing short wavelengths.  It
is not claimed to be derived from the passive vertical spectrum.  Instead, the
passive disk calculation supplies the physical setting for effective drift,
diffusion, leakage, and feedback coefficients.  The steady response to a
localized ring source is then an exactly solvable fourth-order Green function.
When the
right-decaying spatial roots are complex, the exterior branch is a damped
oscillation and the spacing is fixed by the imaginary part of the root.

This construction has three practical advantages.  First, it can operate
without pre-existing planets and is therefore compatible with rings in very
young disks.  Second, the spacing is tied to explicit complex roots, giving a
falsifiable prediction rather than a fitted numerical pattern.  Third, the
forced spatial spacing is separated from the fastest-growing temporal
wavelength, so localized disk features such as snow lines or dead-zone
transitions can be tested directly.

The Keplerian rotation is retained in the starting transport equation.  For an
axisymmetric ring source, however, azimuthal Fourier decomposition excites only
the $m=0$ component, so the rotation and azimuthal-diffusion terms vanish
identically.  Rotation shears non-axisymmetric plumes, but it does not alter
the ring-averaged passive kernel.  This exact passive result motivates the
one-dimensional radial feedback problem used below.

The paper is organized as follows.  The Model section sets up the passive
transport problem and introduces the radial feedback closure.  The
Results section solves the steady forced Green function and gives a numerical
example.  The Discussion compares the temporal and spatial length scales,
explains the role of the passive kernel, and summarizes the limitations of the
local constant-coefficient model.  Full derivations and verification tests are
given in the Supplemental Material.

{\bf\emph{Model.}}---
The model is defined as a passive transport kernel coupled to a local
scale-selecting feedback closure.  The passive kernel describes how a
ring-averaged tracer is transported in a Keplerian disk with vertical
diffusion.  It sets the background response and motivates effective drift,
diffusion, and leakage coefficients.
The ring spacing is then determined by a separate fourth-order radial feedback
operator acting on the vertically integrated perturbation.  This separation is
part of the model: passive transport sets the environment, while the feedback
Green function selects the observable radial scale.

For the passive part, consider the steady-state concentration
$f(r,\phi,z)$ of a tracer in a thin accretion disk.  The advection--diffusion
equation in cylindrical coordinates is
\begin{equation}
  \mathbf{v}\cdot\nabla f
  =\frac{1}{r}\partial_r\!\left(rD_r\partial_r f\right)
  +\frac{D_\phi}{r^2}\partial_\phi^2 f
  +\partial_z\!\left[D_z(z)\partial_z f\right]
  +S,
  \label{eq:full3d}
\end{equation}
where $v_r=-c/r$, $v_\phi=\Omega_K r=\sqrt{GM/r}$, $v_z=0$, and $D_z(z)>0$ is a
general vertical diffusivity profile.  For a ring source at $r=r_0$, $z=z_0$,
$S=\tilde Q\,\delta(r-r_0)\delta(z-z_0)$, only the azimuthal $m=0$ component
is excited.  The rotation and azimuthal diffusion terms therefore vanish, and
the passive axisymmetric equation becomes
\begin{equation}
  \begin{split}
  &D_r\!\left(\partial_r^2+\frac{1}{r}\partial_r\right)\!f
  +\frac{c}{r}\partial_r f
  +\partial_z\!\left[D_z(z)\partial_z f\right]\\
  &\qquad=-\tilde Q\,\delta(r-r_0)\delta(z-z_0).
  \end{split}
  \label{eq:pde}
\end{equation}

We impose midplane symmetry $\partial_z f|_{z=0}=0$ and escape
$f|_{z=L}=0$.  Writing $f=\sum_n a_n(r)Z_n(z)$ gives
\begin{equation}
  \frac{d}{dz}\!\left[D_z(z)\frac{dZ_n}{dz}\right]+\lambda_n Z_n=0,
  \quad Z_n'(0)=0,\quad Z_n(L)=0,
  \label{eq:SL}
\end{equation}
and
\begin{equation}
  D_r a_n''+\frac{D_r+c}{r}a_n'-\lambda_n a_n
  =-Q_n\delta(r-r_0).
  \label{eq:radial}
\end{equation}
The bounded exterior radial solution is
\begin{equation}
  \begin{split}
  a_n^+(r)&=\frac{Q_n}{D_r}\,
  r_0^{1+\nu}I_\nu(\kappa_n r_0)\;
  r^{-\nu}K_\nu(\kappa_n r),\\
  \nu&=\frac{c}{2D_r},\qquad
  \kappa_n=\sqrt{\frac{\lambda_n}{D_r}} .
  \end{split}
  \label{eq:an}
\end{equation}
Thus the passive exterior Green function is
\begin{equation}
  \begin{aligned}
  f^+(r,z)
  &=\frac{\tilde Q}{D_r}\sum_{n=1}^{\infty}
  \frac{Z_n(z_0)Z_n(z)}{\mathcal N_n}\;
  r_0^{1+\nu}I_\nu(\kappa_n r_0)\\
  &\qquad\times r^{-\nu}K_\nu(\kappa_n r),
  \qquad r>r_0 .
  \end{aligned}
  \label{eq:green}
\end{equation}
This expression is exact for the passive ring-source problem and defines the
baseline transport response used below.  The corresponding low modes have long
radial decay lengths, so the passive response is smooth on the scales of
interest.  The subsequent ring spacing is not a consequence of the passive
vertical structure alone; it comes from the separate feedback Green function.
The normalization and radial Green-function checks are given in the
Supplemental Material.

Let
$\Sigma(r,t)$ denote the vertically projected ring-averaged perturbation
amplitude or tracer surface density.  We use
\begin{equation}
  \partial_t\Sigma+\partial_rJ+\lambda\Sigma=S_\Sigma ,
  \label{eq:evol}
\end{equation}
with local feedback flux
\begin{equation}
  J=\left[-U_0+a_1\partial_r\Sigma
  +a_3\partial_r^3\Sigma\right]\Sigma
  -D_\ast\partial_r\Sigma .
  \label{eq:flux}
\end{equation}
Here $U_0$ is a mean drift speed, $D_\ast$ is an effective radial diffusivity,
$\lambda$ is a loss or leakage rate, $a_1$ is a positive gradient feedback, and
$a_3>0$ is the short-wavelength regularization.  These coefficients are
phenomenological in the present minimal model.  A microscopic disk calculation
or simulation would be needed to calibrate them for a particular source.  The
factor multiplying the bracket in Eq.~\eqref{eq:flux} is an explicit local
feedback assumption: drift changes driven by $\partial_r\Sigma$ and
$\partial_r^3\Sigma$ are taken to be stronger where more tracer or solids are
available to participate in the feedback.  Thus the feedback flux is
proportional to the local surface density, while $D_\ast$ represents an
ordinary diffusive flux.

Set $\Sigma=\Sigma_0+\sigma$ and choose $S_\Sigma=\lambda\Sigma_0$ for the
background.  Linearization gives
\begin{equation}
  \begin{split}
  &\partial_t\sigma-U_0\partial_r\sigma
  +A\partial_r^2\sigma+B\partial_r^4\sigma+\lambda\sigma=0,\\
  &A=\Sigma_0a_1-D_\ast,\qquad B=\Sigma_0a_3 .
  \end{split}
  \label{eq:linear}
\end{equation}
The $a_1$ term competes with diffusion.  In the linear equation it contributes
$+\Sigma_0a_1\partial_r^2\sigma$, which is an anti-diffusive gradient feedback
when $A=\Sigma_0a_1-D_\ast>0$.  The fourth derivative is the central
regularizing term: it allows a finite wavelength to be selected while
suppressing arbitrarily short wavelengths.

For a homogeneous perturbation $\sigma\propto\exp(\omega t+ikr)$,
\begin{equation}
  \omega(k)=-\lambda+iU_0k+Ak^2-Bk^4 .
  \label{eq:dispersion}
\end{equation}
The real growth rate is $-\lambda+Ak^2-Bk^4$.  If $A>0$, the fastest-growing
temporal wavenumber is
\begin{equation}
  k_\ast^2=\frac{A}{2B},
  \qquad
  \gamma_{\rm max}=-\lambda+\frac{A^2}{4B}.
  \label{eq:kstar}
\end{equation}
This is a useful stability diagnostic, but it is not the spatial Green
function of a steady ring injector.  The steady-source spacing is computed
from the spatial roots below.

{\bf\emph{Results.}}---
Place a localized ring source at $r=r_0$ and write $x=r-r_0$.  The exact
steady response of the linear feedback operator solves
\begin{equation}
  B\frac{d^4 f}{dx^4}
  +A\frac{d^2 f}{dx^2}
  -U_0\frac{df}{dx}
  +\lambda f
  =\delta(x),
  \label{eq:green_ode}
\end{equation}
with decay as $x\to\pm\infty$.  The homogeneous spatial roots satisfy
\begin{equation}
  Bq^4+Aq^2-U_0q+\lambda=0 .
  \label{eq:spatial_poly}
\end{equation}
The branch outside the injector keeps the two roots with
$\operatorname{Re}q<0$.  If they form a complex pair
$q_\pm=\alpha\pm i\beta$ with $\alpha<0$, then
\begin{equation}
  f_+(x)=C e^{(\alpha+i\beta)x}+C^\ast e^{(\alpha-i\beta)x}
  =2|C|e^{\alpha x}\cos(\beta x+\phi).
  \label{eq:exterior}
\end{equation}
The constants are fixed by continuity of $f$, $f'$, and $f''$ at $x=0$, and by
the delta-function jump
\begin{equation}
  B\left[f'''(0^+)-f'''(0^-)\right]=1 .
  \label{eq:jump}
\end{equation}
The exterior spacing is therefore
\begin{equation}
  \Delta r_{\rm Green}=\frac{2\pi}{\beta},
  \label{eq:green_spacing}
\end{equation}
which is generally distinct from $2\pi/k_\ast$.

\begin{figure}[t]
  \centering
  \includegraphics[width=\columnwidth]{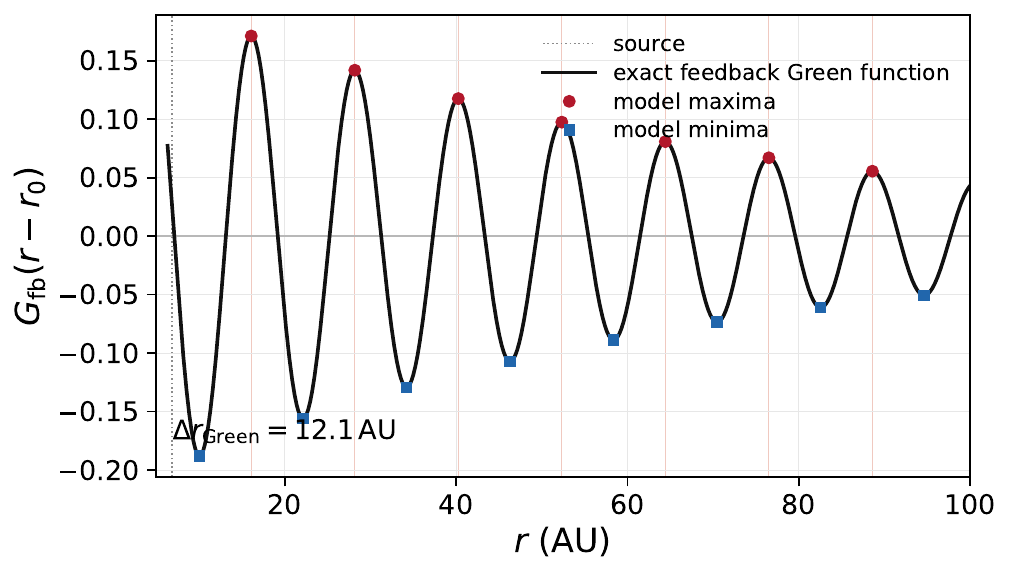}
  \caption{Exact steady Green response of the fourth-order radial feedback
  model, displayed in AU.  The curve is obtained by solving
  Eq.~\eqref{eq:green_ode} with decay on both sides of the source and the jump
  condition Eq.~\eqref{eq:jump}.  Red points mark the analytic maxima of the
  model Green function.  The exterior spacing is
  $\Delta r_{\rm Green}\simeq12.1$~AU.}
  \label{fig:feedback_green}
\end{figure}

For the worked dimensionless example, we use
\begin{equation}
  \begin{gathered}
  U_0=0.3,\qquad \Sigma_0=1,\qquad a_1=5.2,\qquad a_3=0.4,\\
  D_\ast\simeq2.25,\qquad \lambda\simeq2.91 .
  \end{gathered}
  \label{eq:params}
\end{equation}
The numerical values are chosen to give a simple complex-root example; the
Supplemental Material gives the equivalent root-based construction and the
full numerical precision.  The fastest temporal wavelength is about $3.27$ in
these local units, while the steady forced Green function selects the different
spacing
\begin{equation}
  \Delta r_{\rm Green}\simeq2.52 .
  \label{eq:num_spacing}
\end{equation}
Figure~\ref{fig:feedback_green} maps the local coordinate $x$ to AU using an
observationally motivated display scale.  This gives
$\Delta r_{\rm Green}\simeq12.1$~AU in the figure.

The plotted curve is the exact piecewise Green function obtained from
Eq.~\eqref{eq:green_ode}.  The Supplemental Material gives the matching
system, the extrema formula, and numerical checks of the root residuals,
continuity conditions, jump condition, and branch-wise ODE residuals.

{\bf\emph{Discussion.}}---
The DSHARP survey \cite{Andrews2018} found typical ring separations of
20--40~AU in disks around nearby young stars.  HL~Tau shows separations of
$\sim$10--25~AU \cite{ALMA2015}; HD~163296 and AS~209 show multiple rings at
larger radii \cite{Isella2016,Fedele2018,Huang2018}.  In the present model,
matching these physical spacings requires calibrating the dimensionless radial
unit and the feedback coefficients $U_0,D_\ast,\lambda,a_1,a_3$ from disk
microphysics, chemistry, dust dynamics, thermal feedback, non-ideal MHD, or
simulations.  The AU scale in Fig.~\ref{fig:feedback_green} is therefore an
illustrative calibration of the local length unit, not a disk-specific fit of
all coefficients.

The passive calculation remains important because it identifies the transport
kernel on which any ring-forming feedback acts.  The axisymmetric Keplerian
problem separates cleanly, and the passive solution can be written exactly in
terms of vertical Sturm--Liouville modes and modified Bessel functions for a
general vertical diffusivity profile.  In the present construction the passive
Green function supplies the background transport response and the motivation
for effective local coefficients.  The periodic exterior maxima are associated
with the fourth-order feedback operator, not with any special choice of
vertical diffusivity.

The feedback closure should be read as a local amplitude equation rather than a
microscopic disk model.  The factor of $\Sigma$ in Eq.~\eqref{eq:flux} assumes
that gradient-driven corrections to the drift act more strongly where more
tracer or solids are present.  The $a_1$ term is therefore an anti-diffusive
gradient feedback in the linearized problem, while $D_\ast$ is ordinary
diffusion.  The $a_3$ term provides the minimal local regularization that
prevents the model from selecting arbitrarily short wavelengths and allows a
complex decaying spatial branch to dominate.  It should not be interpreted as a
unique statement about the microscopic physics.  Nonlocal transport,
multi-fluid dust--gas coupling, finite thermal or chemical relaxation, and
scale-dependent turbulent stresses could play the same role after reduction to
an effective radial operator.  A disk simulation or a thermochemical dust--gas
model would be needed to compute these coefficients from first principles.

This root-structure viewpoint gives the most reliable route for applying the
mechanism beyond the present minimal closure.  In protoplanetary disks,
dust--gas backreaction and pressure-bump trapping provide direct examples of
transport feedback: local dust concentration changes the gas response and can
further enhance concentration \cite{YoudinGoodman2005,Johansen2009,Pinilla2017}.
Irradiation and shadowing models provide another scale-selective route, because
thermal relaxation and radiative transport regularize a geometric heating
feedback and can generate ring--gap patterns
\cite{WuLithwick2021,Ueda2021,Ziampras2025}.  Dense planetary rings are an
especially clean analogue: in viscous overstability, the effective stress
depends on surface density, extracts energy from Keplerian shear, and produces
axisymmetric wavetrains \cite{Salo2001,Schmidt2001,ReinLatter2013}.  Debris
and circumbinary disks also show ring-like structures
\cite{Hughes2018,Rosenfeld2013}, but they may admit similar effective
descriptions only when their transport can be treated as a continuum feedback
problem; planet-driven resonances or externally imposed cavities should instead
be viewed as forcing mechanisms, not as the feedback mechanism modeled here.

The distinction between $k_\ast$ and $q_+$ is also observationally relevant.
The wavenumber $k_\ast$ belongs to a homogeneous temporal instability.  It
answers which Fourier mode grows fastest if the disk is perturbed everywhere.
The imaginary part of $q_+$ belongs to the steady forced Green function.  It
answers what radial pattern is produced outside a localized injector such as a
snow line, dead-zone transition, or chemically active ring.  The two spacings
need not agree, and in the worked example they differ by about 30 percent.

If the injector is moved to a finite radius such as a snow line, the local
constant-coefficient calculation is unchanged after replacing $x$ by $r-r_0$,
provided the coefficients vary slowly over one spacing.  A large $r_0$ mainly
helps justify this local approximation and keeps the response away from the
inner boundary.  If $U_0$, $A$, $B$, or $\lambda$ vary significantly with
radius, the spatial polynomial must be replaced by a variable-coefficient
calculation; then the oscillations can remain quasi-periodic, but the spacing
will drift with radius.  Likewise, the decay conditions at $x=\pm\infty$ should
be read as the local limit of a finite disk whose boundaries are several decay
lengths from the source.  Inner truncation or outer outflow boundaries would
modify the amplitudes and phases near the boundaries, but not the local
spacing while the complex-root branch remains dominant.

For a point source localized in $\phi$, the $m\ne0$ azimuthal modes are
excited.  Keplerian shear then stretches the plume into an azimuthal filament,
and the radial problem is no longer the simple Bessel kernel.  Azimuthal
diffusion eventually homogenizes the filament, and the long-time ring-averaged
profile is again governed by the $m=0$ component, possibly with an effective
radial diffusivity enhanced by shear dispersion \cite{Taylor1953,Youdin2011}.
For the ring-source problem studied here, this complication is absent.

The model also has clear limitations.  The radial coefficients are taken to be
constant and the feedback is linearized around a uniform background.  Radially
varying coefficients would replace the spatial polynomial by a
variable-coefficient fourth-order equation and would generally produce a slowly
drifting spacing.  Nonlinear saturation, dust settling, and backreaction are
also outside the present calculation.  These effects are needed for ring
contrasts and detailed disk-by-disk modeling, but they are separate from the
scale-selection mechanism isolated here.

{\bf\emph{Conclusion and Outlook.}}---
We have constructed an exactly solvable local model for ring spacing in a
Keplerian disk.  The passive disk supplies a smooth Bessel--Sturm--Liouville
transport kernel: Keplerian rotation drops out for an axisymmetric ring source,
the vertical diffusivity profile gives a spectrum, and the exterior radial
response is a modified-Bessel Green function.  This passive kernel motivates
the effective coefficients used in the radial feedback closure, but the
existence of damped periodic maxima is controlled by the fourth-order feedback
roots rather than by the passive vertical structure.  Because the feedback is
an internal disk process, the mechanism does not require a pre-existing planet
and may operate while planet formation is still underway.

The spacing is selected by the fourth-order radial feedback Green function.  A
steady localized source excites spatial roots of
$Bq^4+Aq^2-U_0q+\lambda=0$, and a complex right-decaying pair gives a damped
oscillatory exterior response.  In the worked dimensionless example,
$\Delta r_{\rm Green}\simeq2.52$, corresponding to about $12$~AU under the
display scaling used in Fig.~\ref{fig:feedback_green}.

The resulting prediction is structural: if a localized disk feature activates
this kind of feedback, the observed spacing should track the imaginary part of
the local spatial Green-function root, not the fastest temporal wavenumber and
not passive Bessel modal crossings.  This is the main advantage over models
that identify rings only after numerical evolution: the proposed spacing can be
computed from the local coefficients before a disk-specific simulation is run.
The next step is to calibrate the feedback coefficients in simulations or
thermochemical dust--gas models and test whether the inferred spatial roots
reproduce ring spacings in individual disks.

\begin{acknowledgments}
During the preparation of this work, the authors used DeepSeekV3 to improve
readability and language.  After using this tool, the authors reviewed and
edited the content as needed and take full responsibility for the content of
the publication.  This work is supported by K. C. Wong Educational
Foundation.
\end{acknowledgments}

\clearpage
\onecolumngrid

\section*{Supplemental Material for\\
``Ring Spacing from a Fourth-Order Radial Feedback Green Function
in a Keplerian Accretion Disk''}

\setcounter{section}{0}
\renewcommand{\thesection}{S\arabic{section}}
\renewcommand{\thefigure}{S\arabic{figure}}
\renewcommand{\theequation}{S\arabic{equation}}
\renewcommand{\thetable}{S\arabic{table}}
\renewcommand{\theHsection}{S\arabic{section}}
\renewcommand{\theHfigure}{S\arabic{figure}}
\renewcommand{\theHequation}{S\arabic{equation}}
\renewcommand{\theHtable}{S\arabic{table}}
\setcounter{figure}{0}
\setcounter{equation}{0}
\setcounter{table}{0}

\section{Overview}
\label{secS:purpose}

This Supplemental Material provides the derivations and numerical checks used
in the main text.  Sections~\ref{secS:closure}--\ref{secS:numerical} derive the
fourth-order radial feedback equation, its homogeneous temporal dispersion
relation, and the exact steady Green function for a localized ring source.
Section~\ref{secS:passive} records the passive vertical transport kernel that
motivates the effective coefficients in the feedback model.  The final section
summarizes the verification tests.

The passive vertical calculation is used as a disk-motivated background kernel.
The finite ring spacing derived here is not a consequence of the vertical
diffusivity profile alone; it is selected by the fourth-order radial feedback
Green function.

The main distinction needed below is between two length scales.  The
homogeneous temporal problem has a fastest-growing Fourier wavenumber
$k_\ast$, whereas a steady localized source excites a spatial Green function.
The ring spacing in the constant-coefficient calculation is determined by the
imaginary part of the right-decaying spatial root $q_+$.

\section{Fourth-Order Radial Feedback Closure}
\label{secS:closure}

Let $\Sigma(r,t)$ be the vertically projected perturbation amplitude or tracer
surface density.  We use the one-dimensional conservation law
\begin{equation}
  \partial_t\Sigma+\partial_rJ+\lambda\Sigma=S_\Sigma ,
  \label{eqS:evol}
\end{equation}
with feedback flux
\begin{equation}
  J=\left[-U_0+a_1\partial_r\Sigma
  +a_3\partial_r^3\Sigma\right]\Sigma
  -D_\ast\partial_r\Sigma .
  \label{eqS:flux}
\end{equation}
Here $U_0$ is a mean drift speed, $D_\ast$ is an effective radial diffusivity,
$\lambda$ is a loss or leakage rate, $a_1$ is the positive gradient feedback,
and $a_3>0$ regularizes short wavelengths.  These are phenomenological
dimensionless parameters in the present local model.  Their physical role is
motivated by the passive vertical transport kernel in Sec.~\ref{secS:passive},
while their numerical calibration is left to disk simulations or microscopic
transport models.  The product
structure in Eq.~\eqref{eqS:flux} is part of the closure: the velocity-like
feedback correction
$a_1\partial_r\Sigma+a_3\partial_r^3\Sigma$ is assumed to act on the local
amount of tracer or solids.  This represents a local feedback whose flux is
stronger where more material participates, while the separate
$-D_\ast\partial_r\Sigma$ term is the ordinary diffusive flux.

Set
\begin{equation}
  \Sigma(r,t)=\Sigma_0+\sigma(r,t),
  \qquad
  S_\Sigma=\lambda\Sigma_0 ,
  \label{eqS:background}
\end{equation}
with constant $\Sigma_0$.  Since
\begin{equation}
  \partial_r\Sigma=\partial_r\sigma,
  \qquad
  \partial_r^3\Sigma=\partial_r^3\sigma,
  \label{eqS:derivs}
\end{equation}
the flux to first order in $\sigma$ is
\begin{equation}
  J=-U_0\Sigma_0-U_0\sigma
  +(\Sigma_0a_1-D_\ast)\partial_r\sigma
  +\Sigma_0a_3\partial_r^3\sigma+O(\sigma^2).
  \label{eqS:linear_flux}
\end{equation}
Substituting into Eq.~\eqref{eqS:evol} gives the linear fourth-order equation
\begin{equation}
  \partial_t\sigma-U_0\partial_r\sigma
  +A\partial_r^2\sigma+B\partial_r^4\sigma
  +\lambda\sigma=0,
  \qquad
  A=\Sigma_0a_1-D_\ast,
  \qquad
  B=\Sigma_0a_3 .
  \label{eqS:linear}
\end{equation}
The sign convention in Eq.~\eqref{eqS:linear} makes
$B\partial_r^4\sigma$ stabilizing at high temporal wavenumber when $B>0$.
The $a_1$ term contributes
$+\Sigma_0a_1\partial_r^2\sigma$ and is therefore anti-diffusive when it
overcomes $D_\ast$; this is the positive gradient feedback that competes with
ordinary radial diffusion.

\section{Homogeneous Temporal Instability}
\label{secS:temporal}

For a Fourier perturbation
\begin{equation}
  \sigma(r,t)=\hat\sigma \exp(\omega t+ikr),
  \label{eqS:fourier}
\end{equation}
Eq.~\eqref{eqS:linear} gives
\begin{equation}
  \omega(k)=-\lambda+iU_0k+Ak^2-Bk^4 .
  \label{eqS:omega}
\end{equation}
The real growth rate is
\begin{equation}
  \gamma(k)=\operatorname{Re}\omega(k)
  =-\lambda+Ak^2-Bk^4 .
  \label{eqS:gamma}
\end{equation}
Finite-wavelength temporal growth requires $A>0$.  The fastest-growing
wavenumber satisfies
\begin{equation}
  \frac{d\gamma}{dk}=2Ak-4Bk^3=0,
  \qquad
  k_\ast^2=\frac{A}{2B},
  \label{eqS:kstar}
\end{equation}
and the maximum growth rate is
\begin{equation}
  \gamma_{\rm max}=-\lambda+\frac{A^2}{4B}.
  \label{eqS:gmax}
\end{equation}
Thus the homogeneous temporal problem is unstable only if
\begin{equation}
  A>0,
  \qquad
  \frac{A^2}{4B}>\lambda .
  \label{eqS:temporal_condition}
\end{equation}

This calculation characterizes temporal growth in a spatially homogeneous
background.  The steady forced response considered next is a spatial Green
function and therefore has, in general, a different characteristic wavelength.

\section{Exact Steady Green Function of the Fourth-Order Operator}
\label{secS:green}

Let $x=r-r_0$ and place a unit steady source at $x=0$.  The exact forced
response solves
\begin{equation}
  Bf^{(4)}(x)+Af''(x)-U_0f'(x)+\lambda f(x)=\delta(x),
  \label{eqS:green_ode}
\end{equation}
with decay as $x\to\pm\infty$.  For $x\ne0$, write $f\propto e^{qx}$.  The
spatial roots obey
\begin{equation}
  P(q)\equiv Bq^4+Aq^2-U_0q+\lambda=0 .
  \label{eqS:poly}
\end{equation}
The right branch $x>0$ uses the two roots with $\operatorname{Re}q<0$, while
the left branch $x<0$ uses the two roots with $\operatorname{Re}q>0$:
\begin{align}
  f_+(x)&=\sum_{j=1}^2 C_j e^{q_j^+x},
  \qquad x>0, \label{eqS:right_branch}\\
  f_-(x)&=\sum_{j=1}^2 L_j e^{q_j^-x},
  \qquad x<0. \label{eqS:left_branch}
\end{align}
Integrating Eq.~\eqref{eqS:green_ode} through $x=0$ gives the jump condition
\begin{equation}
  B\left[f'''(0^+)-f'''(0^-)\right]=1 .
  \label{eqS:jump}
\end{equation}
The lower derivatives are continuous:
\begin{equation}
  f(0^+)=f(0^-),\qquad
  f'(0^+)=f'(0^-),\qquad
  f''(0^+)=f''(0^-).
  \label{eqS:continuity}
\end{equation}
Equations~\eqref{eqS:jump} and \eqref{eqS:continuity} form a $4\times4$
linear system for $C_1,C_2,L_1,L_2$:
\begin{equation}
  \sum_{j=1}^2 C_j(q_j^+)^m-\sum_{j=1}^2 L_j(q_j^-)^m=0,
  \qquad m=0,1,2,
  \label{eqS:matching012}
\end{equation}
\begin{equation}
  B\left[
  \sum_{j=1}^2 C_j(q_j^+)^3
  -\sum_{j=1}^2 L_j(q_j^-)^3
  \right]=1 .
  \label{eqS:matching3}
\end{equation}
Equations~\eqref{eqS:matching012} and \eqref{eqS:matching3} determine the
piecewise Green function used in the main figure.

If the right-decaying roots are a complex pair
\begin{equation}
  q_\pm=\alpha\pm i\beta,\qquad \alpha<0,\qquad \beta>0,
  \label{eqS:qpair}
\end{equation}
then reality of the Green function gives $C_-=C_+^\ast$.  The exterior
response can be written
\begin{equation}
  f_+(x)=C e^{(\alpha+i\beta)x}+C^\ast e^{(\alpha-i\beta)x}
  =2|C|e^{\alpha x}\cos(\beta x+\phi),
  \qquad C=|C|e^{i\phi}.
  \label{eqS:real_branch}
\end{equation}
Stationary points satisfy
\begin{equation}
  f_+'(x)=0
  \quad\Longleftrightarrow\quad
  \tan(\beta x+\phi)=\frac{\alpha}{\beta}.
  \label{eqS:stationary}
\end{equation}
At those points,
\begin{equation}
  f_+''(x)
  =-2|C|e^{\alpha x}
  (\alpha^2+\beta^2)\cos(\beta x+\phi),
  \label{eqS:second_derivative}
\end{equation}
which classifies maxima and minima without numerical differentiation.
Successive maxima are separated by
\begin{equation}
  \Delta r_{\rm Green}=\frac{2\pi}{\beta}.
  \label{eqS:green_spacing}
\end{equation}

\section{Numerical Example}
\label{secS:numerical}

The dimensionless example in the main text uses
\begin{equation}
  U_0=0.3,\qquad
  \Sigma_0=1,\qquad
  a_1=5.2,\qquad
  a_3=0.4,\qquad
  D_\ast=2.254726966240959,\qquad
  \lambda=2.9122262769312304 .
  \label{eqS:params}
\end{equation}
These values are obtained by prescribing a simple set of spatial roots and then
reconstructing the polynomial coefficients.
With
\begin{equation}
  s=0.0743770293222018,\qquad
  \beta=2.491598453644235,\qquad
  \eta=1.079899817194562,
  \label{eqS:root_design}
\end{equation}
we set
\begin{equation}
  P(q)=B\left[(q+s)^2+\beta^2\right]
  \left[(q-s)^2+\eta^2\right],
  \qquad B=0.4 .
  \label{eqS:designed_poly}
\end{equation}
Matching this expression to
$P(q)=Bq^4+Aq^2-U_0q+\lambda$ gives
\begin{equation}
  A=B(\beta^2+\eta^2-2s^2),\qquad
  U_0=2Bs(\beta^2-\eta^2),\qquad
  \lambda=B(s^2+\beta^2)(s^2+\eta^2).
  \label{eqS:designed_coeffs}
\end{equation}
The numerical choices above round this construction, with $a_1=5.2$ and
$D_\ast=a_1-A$ for $\Sigma_0=1$.
Therefore
\begin{equation}
  A=\Sigma_0a_1-D_\ast=2.945273033759041,
  \qquad
  B=\Sigma_0a_3=0.4 .
  \label{eqS:AB}
\end{equation}
The homogeneous temporal calculation gives
\begin{equation}
  k_\ast=1.9187473237,
  \qquad
  \frac{2\pi}{k_\ast}=3.2746288318,
  \qquad
  \gamma_{\rm max}=2.5094195002 .
  \label{eqS:temporal_numbers}
\end{equation}
These numbers distinguish the temporal instability scale from the
steady-source Green-function scale.

For the steady Green function, the spatial polynomial has roots
\begin{equation}
  q=-0.074377029322\pm2.491598453644i,
  \qquad
  q= 0.074377029322\pm1.079899817195i .
  \label{eqS:roots}
\end{equation}
The exterior branch keeps the first pair, and the unit-source coefficient is
\begin{equation}
  C=0.014443607203+0.097809494047i .
  \label{eqS:Cnum}
\end{equation}
Hence
\begin{equation}
  \Delta r_{\rm Green}
  =\frac{2\pi}{2.491598453644}
  =2.5217487585 .
  \label{eqS:spacing_num}
\end{equation}
For $r_0=20$, the first exterior maxima are
\begin{equation}
  r=\{21.9381766519,\;24.4599254104,\;26.9816741689,
  \;29.5034229274,\;32.0251716859,\;34.5469204444,\ldots\}.
  \label{eqS:maxima_num}
\end{equation}
The equal spacing in Eq.~\eqref{eqS:maxima_num} is a consequence of the
constant-coefficient Green function.  If the coefficients vary significantly
with $r$, the same root calculation must be replaced by a variable-coefficient
calculation and the spacing need not remain exactly constant.

\section{Verification Tests}
\label{secS:checks}

The curve in the main text is generated by \texttt{plot\_feedback\_green.py}.
The independent scripts \texttt{verify\_feedback\_green.py} and
\texttt{verify\_feedback\_green.wl} verify the same algebraic construction by
the following tests:
\begin{enumerate}
  \item each reported root satisfies $P(q)=0$;
  \item the two selected right roots have $\operatorname{Re}q<0$ and the two
        selected left roots have $\operatorname{Re}q>0$;
  \item the solved coefficients satisfy continuity of $f$, $f'$, and $f''$ at
        the source;
  \item the third-derivative jump satisfies
        $B[f'''(0^+)-f'''(0^-)]=1$;
  \item the analytic maxima satisfy $f'(x)=0$ and $f''(x)<0$.
\end{enumerate}
These checks verify the algebraic Green function and the extrema reported in
the main text.

\section{Passive Vertical Transport Kernel}
\label{secS:passive}

This section gives the passive vertical transport calculation used as the
background kernel for the local feedback model.

The steady passive tracer equation is
\begin{equation}
  \mathbf v\cdot\nabla f
  =
  \frac{1}{r}\frac{\partial}{\partial r}
  \left(rD_r\frac{\partial f}{\partial r}\right)
  +\frac{D_\phi}{r^2}\frac{\partial^2 f}{\partial\phi^2}
  +\frac{\partial}{\partial z}
  \left[D_z(z)\frac{\partial f}{\partial z}\right]
  +S .
  \label{eqS:full_passive}
\end{equation}
For the Keplerian field
\begin{equation}
  v_r=-\frac{c}{r},\qquad
  v_\phi=\Omega_K(r)r,\qquad
  v_z=0,
  \label{eqS:velocity}
\end{equation}
and an axisymmetric ring source, only the $m=0$ azimuthal Fourier component is
excited.  The rotation term therefore vanishes exactly.  The passive
axisymmetric equation is
\begin{equation}
  D_r\left(\partial_r^2+\frac{1}{r}\partial_r\right)f
  +\frac{c}{r}\partial_r f
  +\partial_z(D_z(z)\partial_z f)
  =-\tilde Q\delta(r-r_0)\delta(z-z_0).
  \label{eqS:axisym_passive}
\end{equation}

Assume $D_z(z)>0$ is sufficiently regular on $0<z<L$, with midplane symmetry
and escape boundary conditions.  The vertical modes solve
\begin{equation}
  \frac{d}{dz}\left(D_z(z)\frac{dZ_n}{dz}\right)+\lambda_nZ_n=0,
  \qquad Z_n'(0)=0,\qquad Z_n(L)=0,
  \label{eqS:SL}
\end{equation}
with eigenvalues $\lambda_n>0$ and orthogonal eigenfunctions $Z_n$.  No special
closed-form vertical profile is required for the radial feedback calculation
below; a specified $D_z(z)$ only fixes the numerical spectrum and the source
projection.
Let
\begin{equation}
  \mathcal N_n=\int_0^L Z_n^2\,dz,
  \qquad
  Q_n=\frac{\tilde Q Z_n(z_0)}{\mathcal N_n}.
  \label{eqS:norm_Q}
\end{equation}
Projecting Eq.~\eqref{eqS:axisym_passive} onto $Z_n$ gives
\begin{equation}
  D_r a_n''+\frac{D_r+c}{r}a_n'-\lambda_n a_n
  =-Q_n\delta(r-r_0).
  \label{eqS:radial_mode}
\end{equation}
Define
\begin{equation}
  \nu=\frac{c}{2D_r},
  \qquad
  \kappa_n=\sqrt{\lambda_n/D_r}.
  \label{eqS:nukappa}
\end{equation}
Writing $a_n=r^{-\nu}y_n$ gives, for $r\ne r_0$,
\begin{equation}
  y_n''+\frac{1}{r}y_n'
  -\left(\kappa_n^2+\frac{\nu^2}{r^2}\right)y_n=0,
  \label{eqS:bessel_reduction}
\end{equation}
which is the modified Bessel equation of order $\nu$ after setting
$x=\kappa_nr$.  Hence $y_n=I_\nu(\kappa_nr)$ or $K_\nu(\kappa_nr)$, and direct
substitution of $r^{-\nu}I_\nu(\kappa_nr)$ or
$r^{-\nu}K_\nu(\kappa_nr)$ into Eq.~\eqref{eqS:radial_mode} gives zero away
from the source.
For $r\ne r_0$, the radial solutions are
\begin{equation}
  a_n(r)=r^{-\nu}
  \left[A_nI_\nu(\kappa_nr)+B_nK_\nu(\kappa_nr)\right].
  \label{eqS:bessel_solution}
\end{equation}
Regularity at the origin selects $I_\nu$ inside the source radius, and decay
at infinity selects $K_\nu$ outside.  The exterior passive Green function is
\begin{equation}
  f^+(r,z)=
  \frac{\tilde Q}{D_r}
  \sum_{n=1}^{\infty}
  \frac{Z_n(z_0)Z_n(z)}{\mathcal N_n}
  r_0^{1+\nu}I_\nu(\kappa_nr_0)
  r^{-\nu}K_\nu(\kappa_nr),
  \qquad r>r_0 .
  \label{eqS:passive_green}
\end{equation}
The low-$\lambda_n$ modes have the longest radial decay lengths and dominate
the complete positive passive response.  In the model of the main text, this
passive Green function supplies the baseline transport response and coefficient
motivation, while the finite spacing is selected by the fourth-order feedback
Green function in Eq.~\eqref{eqS:green_ode}.

\section{Summary of Verified Relations}
\label{secS:summary}

The verification scripts check the following relations:
\begin{enumerate}
  \item a radial feedback closure produces the fourth-order linear operator
        in Eq.~\eqref{eqS:linear};
  \item the homogeneous temporal problem has
        $k_\ast^2=A/(2B)$ and
        $\gamma_{\rm max}=-\lambda+A^2/(4B)$;
  \item the steady ring-source problem is the fourth-order Green equation
        in Eq.~\eqref{eqS:green_ode}, with continuity of
        $f$, $f'$, and $f''$ and the third-derivative jump
        in Eq.~\eqref{eqS:jump};
  \item for a complex right-decaying root pair, the exterior spacing is
        $\Delta r_{\rm Green}=2\pi/|\operatorname{Im}q_+|$;
  \item passive vertical diffusion supplies a background transport kernel
        and motivates the effective coefficients in the local feedback model;
        it is not the mathematical source of the feedback spacing.
\end{enumerate}

\end{document}